\documentclass[pdflatex,sn-nature]{sn-jnl}

\usepackage{graphicx} 
\usepackage{lineno}
\usepackage{booktabs}
\usepackage{tabularx}
\usepackage{array}
\usepackage{makecell}
\usepackage{multicol}

\usepackage{amsmath}
\usepackage{graphicx}%
\usepackage{multirow}%
\usepackage{amsmath,amssymb,amsfonts}%
\usepackage{amsthm}%
\usepackage{mathrsfs}%
\usepackage[title]{appendix}%
\usepackage{xcolor}%
\usepackage{textcomp}%
\usepackage{manyfoot}%
\usepackage{booktabs}%
\usepackage{algorithm}%
\usepackage{algorithmicx}%
\usepackage{algpseudocode}%
\usepackage{listings}%

\theoremstyle{thmstyleone}%
\theoremstyle{thmstyletwo}%

\theoremstyle{thmstylethree}%

\begin{document}

\title[The potential and viability of V2G for California BEV drivers]{The potential and viability of V2G for California BEV drivers}


\author*[1]{\fnm{Clement} \sur{Wong}}\email{clemwong@umich.edu}

\author[2]{\fnm{Amalie} \sur{Trewartha}}\email{amalie.trewartha@tri.global}
\equalcont{These authors contributed equally to this work.}

\author[2]{\fnm{Steven B.} \sur{Torrisi}}\email{steven.torrisi@tri.global}
\equalcont{These authors contributed equally to this work.}

\author[3]{\fnm{Alexandre} \sur{Filipowicz}}\email{alsfilip@gmail.com}
\equalcont{These authors contributed equally to this work.}

\affil*[1]{\orgdiv{Mechanical Engineering}, \orgname{University of Michigan}, \orgaddress{\street{2350 Hayward}, \city{Ann Arbor}, \postcode{48109}, \state{MI}, \country{USA}}}

\affil[2]{\orgdiv{Energy \& Materials}, \orgname{Toyota Research Institute}, \orgaddress{\street{4440 El Camino Real}, \city{Los Altos}, \postcode{94022}, \state{CA}, \country{USA}}}

\affil[3]{\orgdiv{Human-Centered AI}, \orgname{Toyota Research Institute}, \orgaddress{\street{4440 El Camino Real}, \city{Los Altos}, \postcode{94022}, \state{CA}, \country{USA}}}


\abstract{Vehicle-to-Grid (V2G) adoption is hindered by uncertainties regarding its effects on battery lifetime and vehicle usability. These uncertainties are compounded by limited insight into real-world vehicle usage. Here, we leverage real-world Californian BEV usage data to design and evaluate a user-centric V2G strategy. We identified four clustered driver profiles for V2G assessment, ranging from ``Daily Chargers'' to ``Public Chargers''. We show that V2G participation is most feasible for ``Daily Chargers,'' and that the effects on battery lifetime depend on calendar aging sensitivity. For batteries with low sensitivity, V2G participation increases capacity loss for all drivers. However, for batteries with high sensitivity, V2G participation can lead to negligible changes in capacity or even improved capacity retention, particularly for drivers who tend to keep their batteries at high states of charge. Our findings enable stakeholders to better assess the potential and viability of V2G adoption.}




\maketitle

\section{Introduction}\label{sec1}

Vehicle-to-Grid (V2G) enables the high-capacity batteries in battery electric vehicles (BEVs) to function as stationary energy storage systems that supply and store energy to and from the grid~\cite{kempton1997electric}. V2G has garnered substantial interest as a means of improving grid stability and supporting the integration of renewable energy sources~\cite{xu2023electric, lenka2021grid, yang_2024}. Beyond enhancing the grid, V2G offers a potential revenue stream for BEV owners, as utilities would compensate them for every unit of energy returned to the grid~\cite{kempton2005vehicle, zheng2023modeling}. Furthermore, from the perspective of original equipment manufacturers (OEMs), V2G capabilities can potentially enhance the market appeal of vehicles \cite{sovacool2018neglected, rao_2025_review}.

However, despite these compelling benefits, the widespread adoption of V2G has been hindered by uncertainties over V2G's impact on battery lifetime and vehicle usability~\cite{goncearuc2024barriers, Monica_AutoUI}. Vehicle owners are concerned that adopting V2G will accelerate battery degradation, which can diminish their BEV’s resale value, and leave them with insufficient range for daily driving needs~\cite{Monica_AutoUI, GESKE_driver_concerns_V2G, huang2021electric_EV_driver_concerns}. Utilities may be hesitant to adopt V2G, as consumer concerns with V2G may lead to uncertain participation rates that make it difficult to rely on V2G for grid reinforcement~\cite{goncearuc2024barriers} and predict potential additional loads onto the grid~\cite{li2021coordinating}. Additionally, vehicle OEMs may fear that BEV owners' concerns may negatively impact BEV sales and harm brand loyalty~\cite{goncearuc2024barriers}. Therefore, it is important to have a good understanding of the impact that V2G could have on battery lifetime and vehicle usability in order to assess the potential and viability of V2G for all stakeholders involved.


A number of studies have examined the effects of V2G on battery degradation; however, the results from these studies are often conflicting. Some analyses show that V2G accelerates battery degradation~\cite{Thingvad2021, Dubarry2017, wang2016_V2G_impact_degradation, Bishop2013}, whereas others report that the impact can be negligible or even improve battery longevity~\cite{baure_V2G_neglible_impact_V2G, Uddin2018, Wei2022,Gong2024}. These divergent outcomes stem from V2G's disparate effects on distinct aging mechanisms: \emph{cycle aging} and \emph{calendar aging}~\cite{Hamid_V2G}. \emph{Cycle aging} is the degradation caused by charging and discharging the battery. \emph{Calendar aging} is passive degradation when the battery is not used and is accelerated by time spent at high State of Charge (SOC) and high temperatures~\cite{gasper2023degradation}. V2G participation intuitively increases cycle aging by introducing additional charge/discharge cycles. However, V2G can also reduce calendar aging by allowing the battery to spend more of its time resting at a lower SOC~\cite{Hamid_V2G, Uddin2018}. Thus, V2G's net impact on battery health depends on the tradeoffs between increased cycle aging and potentially decreased calendar aging. These changes depend on vehicle usage behaviors (i.e., driving, charging and V2G participation patterns)~\cite{uddin2017, Wei2022}, battery design ~\cite{petit2016, Kim2022, Leippi2022}, and environmental temperature~\cite{Hamid_V2G, Guo2019, Thingvad2021}. Variations across these factors create many possible scenarios, which explain the conflicting findings in the literature and make it challenging to comprehensively assess V2G's impact on battery degradation.

A major limitation of existing research is the restricted understanding of real-world BEV usage behavior~\cite{yang_2024, Methodology2023, shariff2019state}. Most existing studies make simplifying assumptions about BEV usage behavior, such as drivers charging their vehicles to 100\% nightly~\cite{Uddin2018}, that do not capture the wide variability of real-world BEV usage. 

In this study, we leverage real-world usage data from BEVs in California to resolve the key uncertainties hindering V2G adoption. This work is organized as follows:
\begin{enumerate}
    \item We design a user-centric V2G strategy optimized to balance driver convenience, minimal battery degradation, and maximum utility compensation~(strategy in Section \ref{subsec:V2G_strategy}). This strategy is informed by real-world vehicle usage behavior, battery degradation models, and proposed V2G compensation rates from utilities ~\cite{SCE_V2G_pilot_program}.
    \item We identify four BEV usage profiles that capture a range of real-world Californian BEV usage behaviors critical for V2G assessment. Identifying these profiles enables us to estimate interactions between different vehicle usage behaviors and V2G participation on battery degradation~(results in Section \ref{subsec:vehicle_behaviors}).
    \item We quantify battery degradation under each vehicle’s usage without V2G, demonstrating how vehicle usage affects the degradation of batteries with different sensitivities to calendar aging versus cycle aging~(results in Section \ref{subsec:no_V2G_simulations}).
    \item We simulate the integration of our user-centric V2G strategy into each vehicle's usage, quantifying the potential participation rates and the resulting impact on battery health for each vehicle usage behavior profile. Our analysis highlights that the effect of V2G participation on battery degradation depends on the battery design’s sensitivity to calendar aging and the extent to which a driver’s behavior amplifies calendar aging~(results in Section \ref{subsec:V2G_simulations}).

\end{enumerate}

By demonstrating the participation rates and battery health effects of a user-centric V2G strategy across diverse driver behaviors and battery designs, this study provides an outlook on the potential and viability of V2G for BEV drivers, utilities, and OEMs.

\section{Results}\label{sec2}

\subsection{V2G Strategy}
\label{subsec:V2G_strategy}

Our V2G strategy simulated in this study was designed to be user-centric and was informed by real-world usage behaviors in our vehicle dataset (Section \ref{subsec:vehicle_dataset}), battery degradation models (Section \ref{subsec:battery_model}), and a V2G pilot program proposed by Southern California Edison (SCE)~\cite{SCE_V2G_pilot_program}. The strategy was optimized to balance three key objectives: minimize the need for BEV owners to alter their habits, reduce additional battery degradation from incorporating V2G, and maximize monetary compensation from utilities. 
While it is a precondition for V2G, we also assume the availability of a bidirectional charger that can output a high wattage back to the grid where the user is able to charge. In the SCE proposed pilot program, scenarios are considered with bidirectional AC home charging at speeds of 6.6 kW, 9.6 kW, and 19.2 kW. For our simulations we chose their middle 9.6 kW speed.


\begin{figure}[h!]
    \centering
    \includegraphics[width=0.75\linewidth]{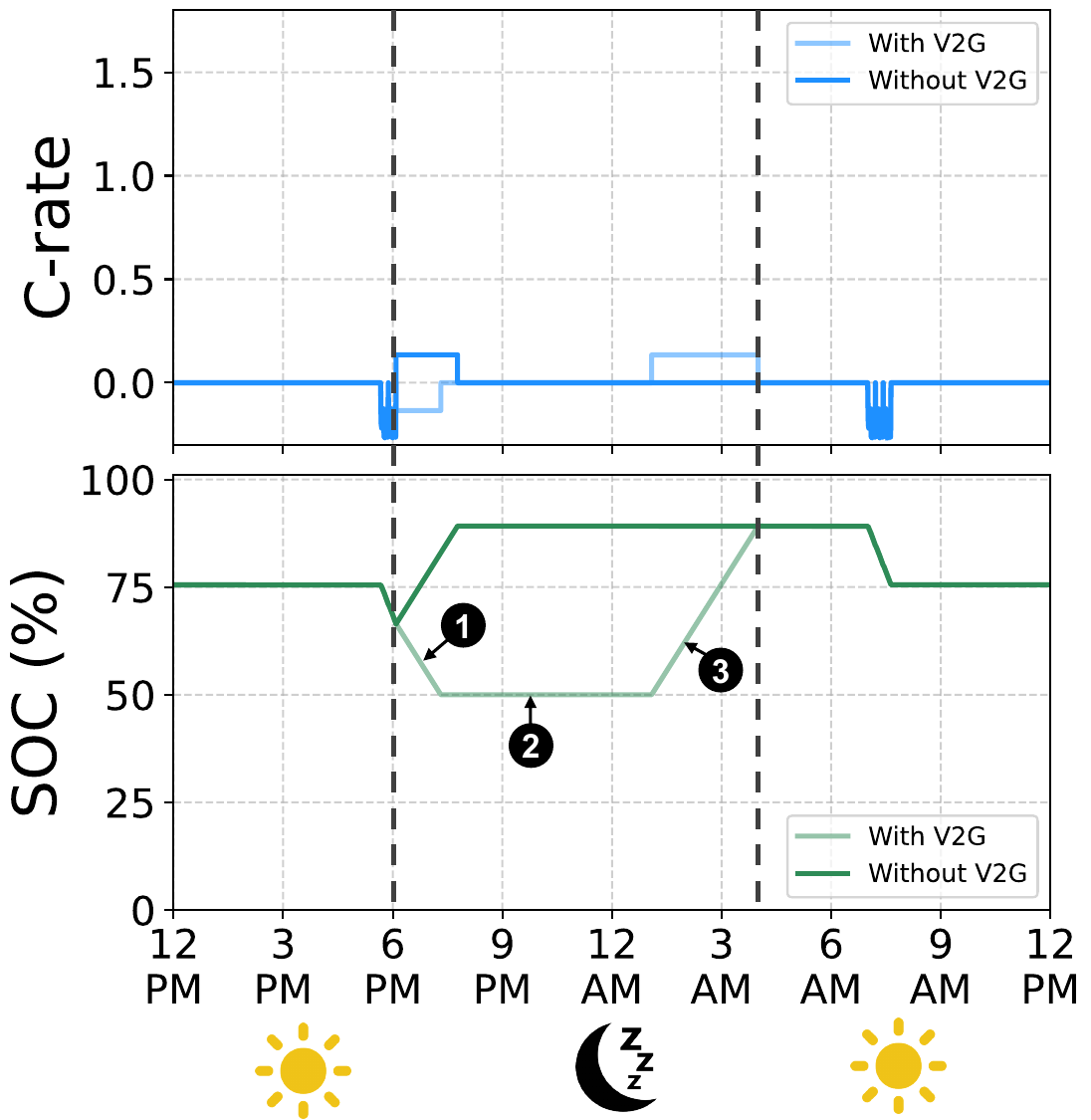}
    \caption{Example of a driver's usage profile, with our V2G strategy compared to it in a conventional charging scenario where the battery begins charging immediately upon plug-in. The V2G strategy consists of 1) discharging to the grid between 6 PM and 9 PM, stopping either 9 PM or when the battery reaches 50\% SOC, whichever occurs first, 2) battery rest period, 3) charging to the driver’s desired SOC by 4 AM.}
    \label{fig:V2G_strategy}
\end{figure}

Figure \ref{fig:V2G_strategy} illustrates our V2G strategy integrated to a driver's usage profile on an example day, comparing it to a conventional charging scenario where the battery begins charging immediately upon plug-in. In the V2G strategy, the vehicle first discharges its battery to the grid at a constant rate of 9.6 kW during the 6 PM to 9 PM peak window. This discharge process stops at 9 PM or if the battery SOC reaches 50\%, whichever occurs first. Following this discharge period, the battery rests before charging to meet the driver’s desired SOC by 4 AM.

The parameters of our V2G strategy were deliberately chosen to meet our key objectives. First, to ensure user convenience, the strategy is designed around overnight charging, a common practice identified in our vehicle usage data. An overnight V2G protocol should not be disruptive to BEV owners, as it operates while vehicles are typically idle. The strategy guarantees the vehicle is charged to the owner's desired level by 4 AM to accommodate the majority of morning commute schedules. To ensure driver utility and prevent range anxiety, we set a 50\% SOC floor based on survey results indicating this level as the minimum charge BEV owners would accept while participating in V2G (see Appendix \ref{sec: survey}). Second, to minimize battery degradation associated with V2G participation, the V2G strategy delays charging after discharge. In comparison to conventional overnight charging where the vehicle charges immediately upon plug-in, this approach enables the battery to spend more time at mid-SOC rather than high SOC, which reduces calendar aging and helps offset the degradation from V2G cycling. Finally, to maximize monetary benefits for BEV owners, we have discharging occur between 6 PM and 9 PM, a peak demand period when Southern California Edison offers the highest compensation for V2G participation~\cite{SCE_V2G_pilot_program}.

\subsection{Identification of Key Californian Vehicle Usage Behavior Profiles for V2G Assessment}
\label{subsec:vehicle_behaviors}

\begin{figure}[h!]
    \centering
    \includegraphics[width=\linewidth]{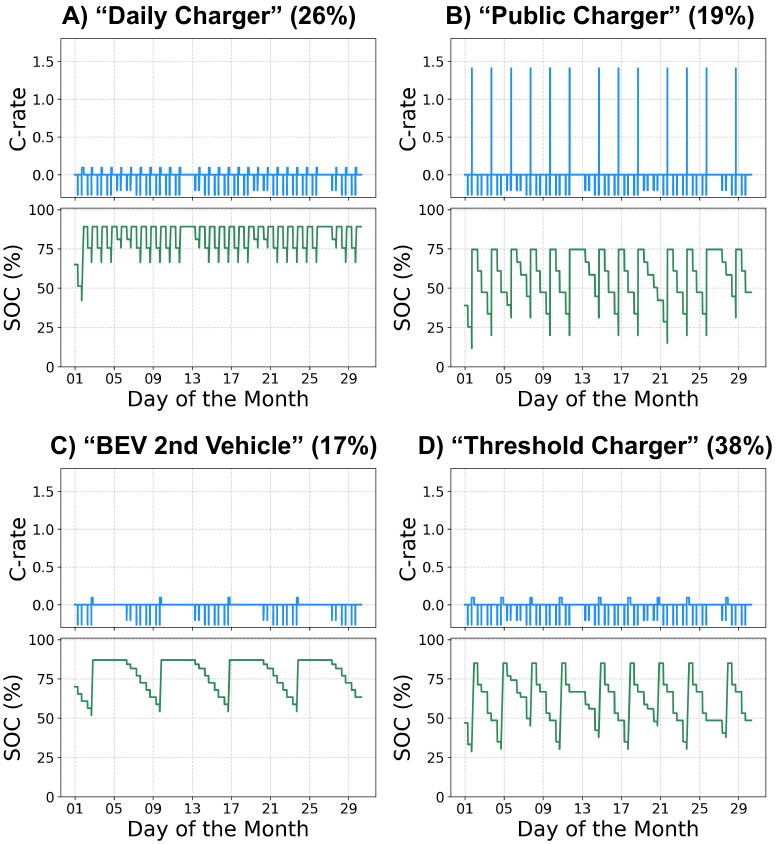}
    \caption{Example month of usage of four key profiles of Californian vehicle usage behavior critical for V2G assessment. Displayed driver profiles are synthetic constructs generated from the average values of each cluster’s behavioral features; individual vehicle usage data is not displayed to ensure driver confidentiality.}
    \label{fig:vehicle_clusters}
\end{figure}

We identified four main profiles of Californian vehicle usage behavior, each with unique implications for battery degradation and V2G viability. Figure~\ref{fig:vehicle_clusters} shows a representative month of usage for vehicles exemplifying each of the four Californian vehicle usage behavior profiles. Table \ref{table:vehicle_clusters} summarizes key behavioral features that distinguish the four identified profiles, including metrics that affect both battery degradation (e.g., driving frequency, charging frequency, SOC before and after charging, and fast charging usage) and V2G integration feasibility (e.g., home charging access).
The first profile, termed the ``Daily Charger'' (26\% of the study sample, Figure \ref{fig:vehicle_clusters}A), represents drivers who consistently charge their vehicles every night after their daily usage. Drivers classified within the ``Daily Charger'' profile tend to use their vehicle frequently, charge frequently, and keep their vehicle at a relatively high average SOC (see Table \ref{table:vehicle_clusters}). The second profile, termed the ``Public Charger'' (19\% of the study sample, Figure \ref{fig:vehicle_clusters}B), represents drivers who also use their vehicle frequently but likely lack home charging access and rely primarily on DC fast charging from public chargers. Drivers classified within the ``Public Charger'' profile charge infrequently and typically charge only up to 80\% SOC, a common upper limit for fast charging at public chargers. The third profile, termed the ``BEV 2nd Vehicle'' (17\% of the study sample, Figure \ref{fig:vehicle_clusters}C), represents drivers who drive their BEV considerably less frequently than drivers in the other profiles (see Table \ref{table:vehicle_clusters}), potentially using their BEV as a secondary vehicle (although this could not be confirmed from our data). Drivers classified within the ``BEV 2nd Vehicle'' profile also charge less frequently, generally opting to charge their BEV when the SOC hits a lower threshold (average SOC before charging = 54.3\%). The fourth profile, termed the ``Threshold Charger'' (38\% of the sample, Figure \ref{fig:vehicle_clusters}D), represents drivers who use their BEVs daily but charge only when necessary rather than every day (see Table \ref{table:vehicle_clusters}).

\begin{table}[h!]
\centering
\caption{Key characteristics of Californian vehicle usage behavior influencing battery degradation and V2G integration feasibility}
\label{table:vehicle_clusters}
\renewcommand{\arraystretch}{1.3}
\begin{tabularx}{\textwidth}{>{\raggedright\arraybackslash}Xcccc}
\toprule
\textbf{} 
& \makecell{\textbf{Daily}\\\textbf{Charger}\\(26\%)} 
& \makecell{\textbf{Public}\\\textbf{Charger}\\(19\%)} 
& \makecell{\textbf{BEV 2nd}\\\textbf{Vehicle}\\(17\%)} 
& \makecell{\textbf{Threshold}\\\textbf{Charger}\\(38\%)} \\
\midrule
Mean number of driving days per week & 6.43 & 6.41 & 4.74 & 6.44 \\
Mean number of charging events per week & 6.11 & 3.34 & 1.87 & 2.57 \\ 
Home Charging Access & Yes & No & Yes & Yes \\
Proportion of Fast Charging Events   & 4.01\% & 74.1\% & 13.8\% & 11.7\% \\
Mean SOC before Charge & 64.7\% & 38.1\% & 54.3\% & 46.1\% \\
Mean SOC after Charge & 89.2\% & 74.7\% & 87.0\% & 85.0\% \\
\bottomrule
\end{tabularx}
\end{table}

Having identified clusters of vehicle usage behaviors, we next examined how variations in vehicle usage behaviors influence battery degradation across different battery designs (Section \ref{subsec:no_V2G_simulations}). We then incorporated our V2G strategy into these vehicle usage behaviors to assess both the feasibility of V2G participation and its impact on battery degradation (Section \ref{subsec:V2G_simulations}).

\subsection{Simulations of Battery Degradation under Californian Vehicle Usage}
\label{subsec:no_V2G_simulations}

\begin{figure} [h!]
    \centering
    \includegraphics[width=\linewidth]{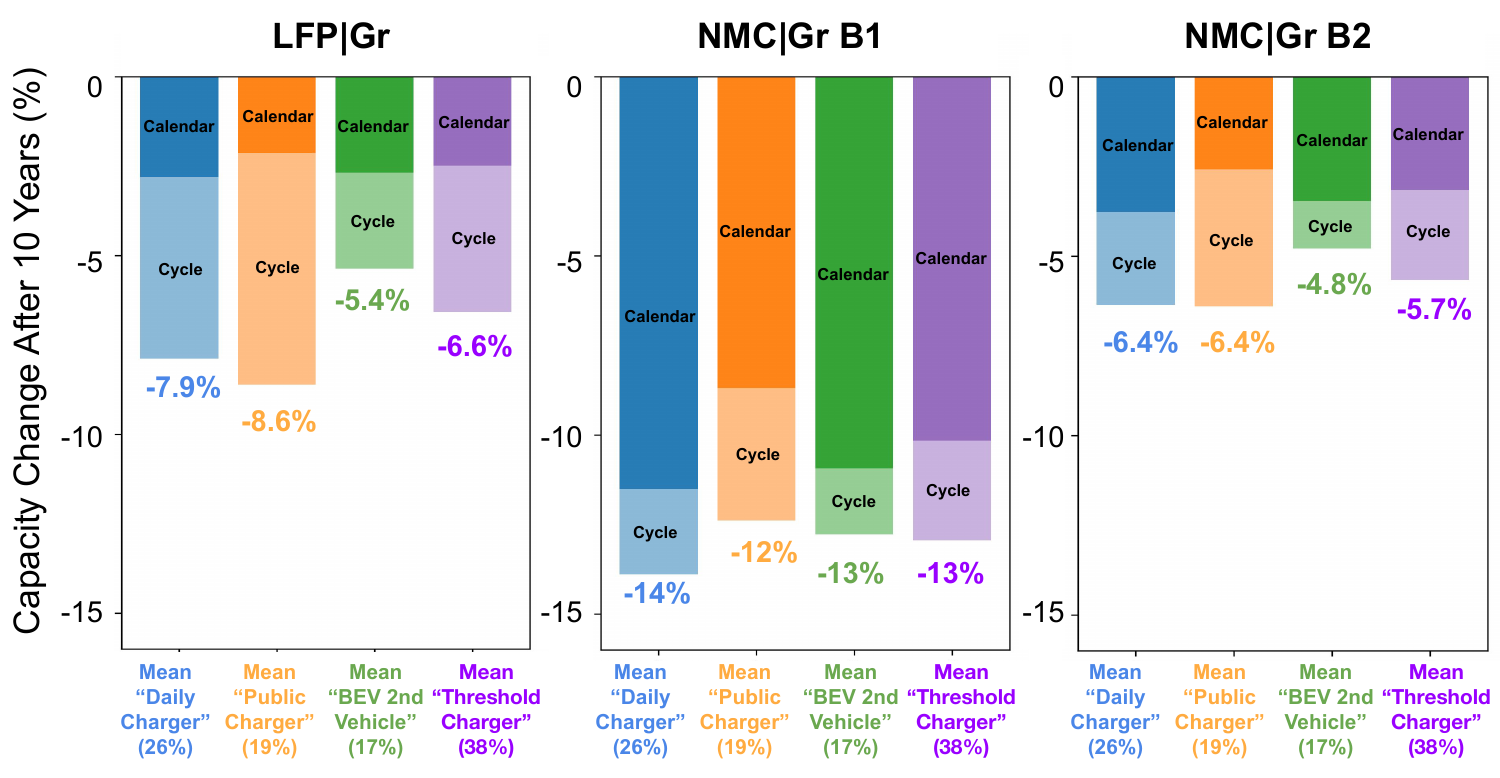}
    \caption{Simulated 10-year capacity loss for three battery designs under the usage of the four Californian vehicle usage behavior profiles. See Appendix Figure \ref{fig:battery_degradation_vehicle_clusters_with_error} for results with 95\% confidence intervals.}
    \label{fig:battery_degradation_vehicle_clusters}
\end{figure}

Figure \ref{fig:battery_degradation_vehicle_clusters} illustrates the effects of Californian vehicle usage behaviors on battery degradation, showing the estimated 10-year capacity loss of various battery designs under the usage behaviors of each California vehicle usage behavior cluster. Each plot highlights the extent to which degradation is attributed to cycle aging and calendar aging, which provides insight into why a battery’s degradation would be more sensitive to certain behaviors than others.

\begin{figure} [h!]
    \centering
    \includegraphics[width=0.75\linewidth]{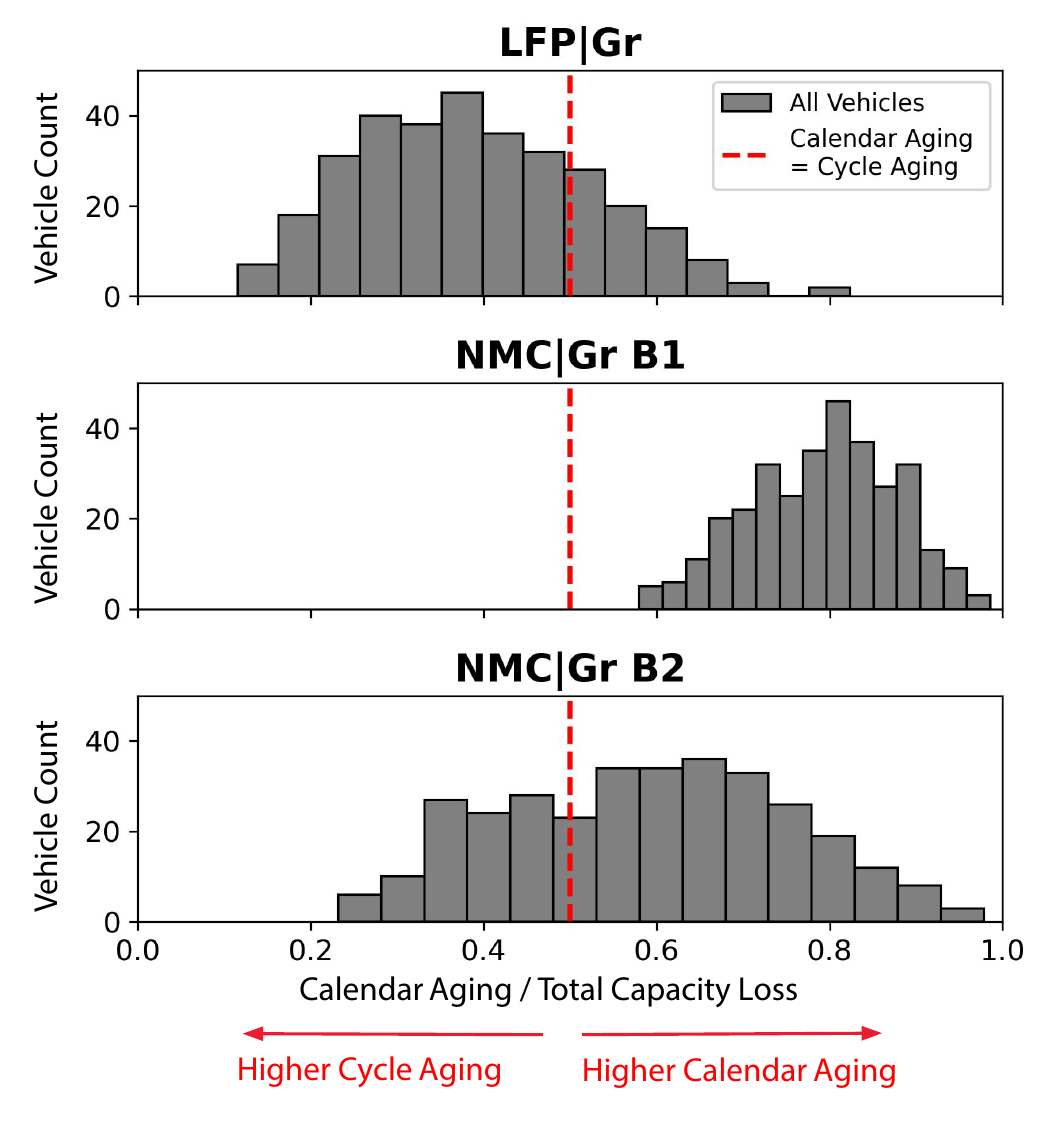}
    \caption{Proportion of total capacity loss after 10 years attributed to calendar aging for all California vehicles, shown separately for each battery design. Capacity loss due to cycle aging is the difference between total capacity loss and capacity loss from calendar aging. }
    \label{fig:calendar_aging_proportion}
\end{figure}

Degradation of the $\mathrm{LFP}\mid\mathrm{Gr}$ battery design is primarily driven by cycle aging (see Appendix Table \ref{tab:calendar_vs_cycle} for Wilcoxon signed-rank test results), which accounts for an average 60.5\% total capacity loss across all vehicle usage behaviors (95\% CI: 59.0\%–62.0\%; Figure \ref{fig:calendar_aging_proportion}). As a result, the $\mathrm{LFP}\mid\mathrm{Gr}$ design’s degradation is primarily driven by behaviors that accelerate cycle aging, such as frequent fast charging and high cycling frequency. Drivers classified within the ``Public Charger'' profile exhibited substantially higher overall capacity loss than those in other behavior groups, primarily due to their more frequent use of fast charging (Figure \ref{fig:battery_degradation_vehicle_clusters}; Kruskal-Wallis: $\chi^2$ = 80.8, p = 2.10 e-17; Mann-Whitney U test comparing degradation from ``Daily Charger'' to all other groups: all ps $<$ 0.05, see Appendix Table \ref{tab:LFP|Gr_public_charger}).

Conversely, degradation of the $\mathrm{NMC}\mid\mathrm{Gr~B1}$ battery design is highly dominated by calendar aging (see Appendix Table \ref{tab:calendar_vs_cycle} for Wilcoxon signed-rank test results). Across all vehicle usage behaviors, calendar aging accounts for on average 79.4\% of total degradation (95\% CI: 78.5\% - 80.3\%; Figure \ref{fig:calendar_aging_proportion}). As a result, the $\mathrm{NMC}\mid\mathrm{Gr~B1}$ design’s degradation is primarily driven by behaviors that increase calendar aging, such as leaving the battery at high SOC for extended periods.  Drivers classified within the ``Daily Charger'' profile tend to leave their batteries at high SOC more frequently than other groups and, as a result, exhibit significantly higher capacity loss (Figure \ref{fig:battery_degradation_vehicle_clusters}; Kruskal-Wallis test e.g., $\chi^2$ =42.9, p=2.64 e-9;  Mann-Whitney U test comparing degradation from ``Daily Charger'' group to all other groups: all ps $<$.05, see Appendix Table \ref{tab:NMC|Grb1_daily_charger}). 

Similarly, the $\mathrm{NMC}\mid\mathrm{Gr~B2}$ battery design is also primarily dominated by calendar aging, though to a lesser extent than $\mathrm{NMC}\mid\mathrm{Gr~B1}$ (see Appendix Table \ref{tab:calendar_vs_cycle} for Wilcoxon signed-rank test results). Across all vehicle usage behaviors, calendar aging accounts for an average of 59.1\% of total degradation (95\% CI: 57.3\%–60.9\%; Figure \ref{fig:calendar_aging_proportion}). Because degradation is more evenly distributed between calendar and cycle aging, the design's capacity loss is not primarily driven by behaviors that increase calendar aging, nor is it primarily driven by behaviors that increase cycle aging. Drivers classified within the ``Public Charger'' and ``Daily Charger'' profiles experienced statistically equivalent levels of degradation (Mann-Whitney U tests: $p = 0.36$, see Appendix Table \ref{tab:NMC|Grb2_public_charger}).

Our simulations demonstrate that the effects of vehicle usage behavior on battery degradation depend on the battery design's sensitivity to calendar aging versus cycle aging. Given this dependency, the effects of integrating V2G into vehicle usage on battery degradation presented in Section \ref{subsec:V2G_simulations} should be understood with consideration of the battery's dominant aging mechanism.

\subsection{Simulations of Incorporating V2G Strategy to California Vehicle Usage}
\label{subsec:V2G_simulations}

We simulated the integration of our V2G strategy (see Section \ref{subsec:V2G_strategy}) to each vehicle usage in our dataset by replacing every overnight charging event where a vehicle was plugged-in and idle between 9 PM and 4 AM with our V2G strategy. 

\begin{figure} [h!]
    \centering
    \includegraphics[width=.8\linewidth]{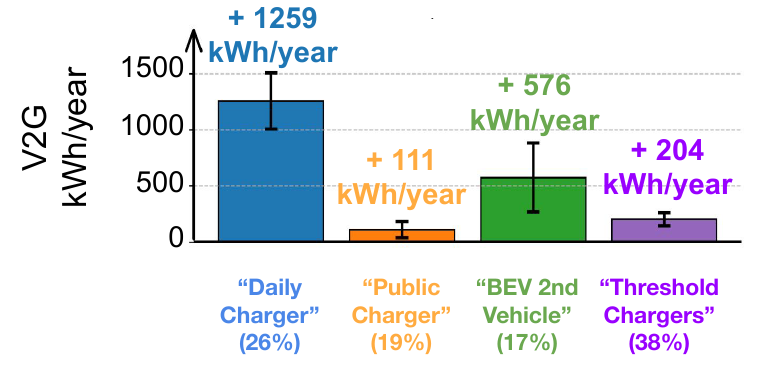}
    \caption{V2G Participation with integration of our V2G strategy (Section \ref{subsec:V2G_strategy}) without vehicle usage behavior change. Bar plots show the population average for each vehicle usage profile; error bars represent the 95\% confidence interval.
}
    \label{fig:participation_rates}
\end{figure}

Figure \ref{fig:participation_rates} shows the average annual energy that drivers classified within each vehicle usage profile could supply to the grid under this strategy, without any changes to their existing behavior. Drivers classified within the ``Daily Charger'' profile exhibited the highest potential, providing an average of 1259 kWh/yr (95\% CI: 1008–1510 kWh/yr). This contribution could translate to an annual revenue of \$406.64/yr (95\% CI: \$325.58 - \$487.73/yr)~\footnote{Revenue calculated based on proposed Southern California Edison's summer rates \cite{SCE_V2G_pilot_program}: a V2G discharge credit of \$0.58/kWh (on-peak, 6 p.m.–9 p.m.) and a charging cost of \$0.257/kWh (off-peak, 9 p.m.–8 a.m.).}. Conversely, drivers classified within the ``Public Charger'' profile offered the least potential, averaging just 111 kWh/yr (95\% CI: 40–182 kWh/yr), due to their infrequent overnight charging. Drivers classified within the ``BEV 2nd Vehicle'' and ``Threshold Charger'' profiles provided moderate contributions, with average annual energy provisions of 576 kWh/yr (95\% CI: 269–882 kWh/yr) and 204 kWh/yr (95\% CI: 144–263 kWh/yr), respectively. Drivers classified within the ``BEV 2nd Vehicle'' and ``Threshold Charger'' profiles could attain participation levels comparable to the ``Daily Charger'' group if they plugged in their vehicles overnight more frequently.

\begin{figure}[h!]
    \centering
    \includegraphics[width=1\linewidth]{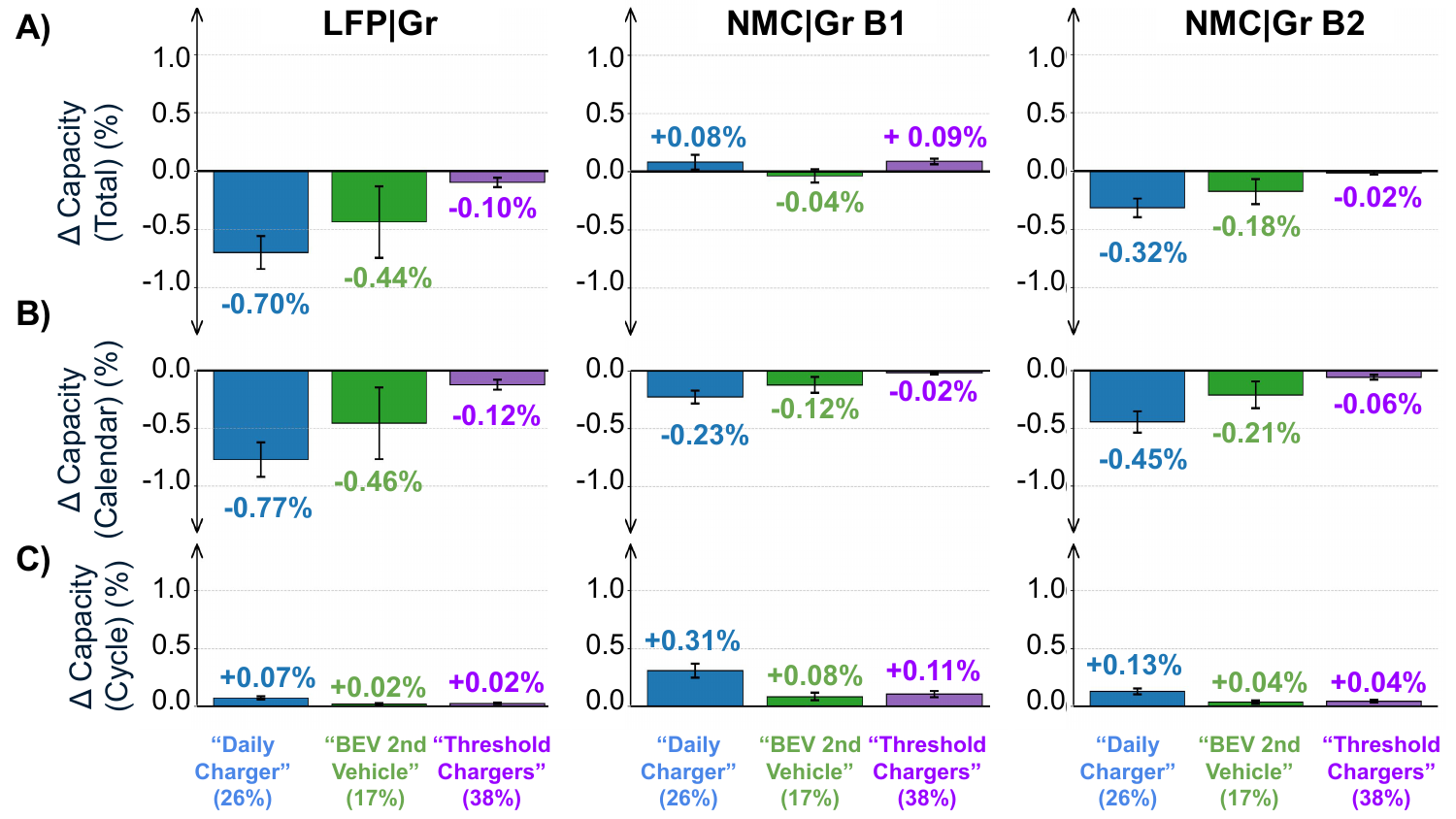}
    \caption{Simulations of the average change in capacity after 10 years with V2G participation. A) Change in total capacity. B) Change in capacity attributed to cycle aging. C) Change in capacity attributed to calendar aging. Error bars correspond to 95\% confidence intervals.
}
    \label{fig:v2g_battery_degradation_vehicle_clusters}
\end{figure}

Figure \ref{fig:v2g_battery_degradation_vehicle_clusters}  illustrates how V2G participation affects degradation across different battery designs for each behavioral cluster. Specifically, it details the change after 10 years in total capacity (Figure \ref{fig:v2g_battery_degradation_vehicle_clusters}A), cycle aging (Figure \ref{fig:v2g_battery_degradation_vehicle_clusters} B), and calendar aging (Figure \ref{fig:v2g_battery_degradation_vehicle_clusters} C) resulting from V2G participation.

With the $\mathrm{LFP}\mid\mathrm{Gr}$ battery design, V2G participation results in greater capacity loss for all vehicle usage profiles (linear regression comparing the influence of V2G kWh/year on total capacity change: all $\beta < -5.18 \times 10^{-4}$, all $p < 0.001$; Appendix Figure \ref{fig:linear_regression_LFP}). The primary reason for the greater capacity loss is the $\mathrm{LFP}\mid\mathrm{Gr}$ battery design’s high sensitivity to cycle aging (Section \ref{subsec:no_V2G_simulations}). The additional cycling induced by V2G participation results in significant cycle aging (linear regression comparing the influence of V2G kWh/year on change in cycle aging for all vehicle usage profiles: all $\beta < -5.93 \times 10^{-4}$, all $p < 0.001$). This increase in cycle aging outweighs any potential reduction in calendar aging from maintaining the battery at a lower SOC during V2G participation, ultimately leading to more capacity loss for all drivers (linear regression comparing the influence of V2G kWh/year on calendar aging for all vehicle usage profiles: all $\beta > 0.23 \times 10^{-4}$, all $p < 0.001$). Figure \ref{fig:v2g_battery_degradation_vehicle_clusters} shows the change in capacity after 10 years with V2G participation by profile, with drivers classified within the ``Daily Charger'' profile experiencing the highest loss due to their greater V2G usage.

In contrast, with the $\mathrm{NMC}\mid\mathrm{Gr~B1}$ battery design, V2G participation does not universally decrease capacity for all vehicle usage profiles; in some cases, it even improves capacity. Unlike the $\mathrm{LFP}\mid\mathrm{Gr}$ battery design, degradation in the $\mathrm{NMC}\mid\mathrm{Gr~B1}$ design is dominated by calendar aging (Section \ref{subsec:no_V2G_simulations}), which is strongly influenced by the battery remaining at high SOC for extended periods. Our V2G participation strategy reduces the amount of time the battery spends at high SOC, thereby significantly reducing calendar aging. This reduction in calendar aging can offset or exceed the additional cycle aging from V2G participation, resulting in a negligible change or even slight improvement in battery capacity. For drivers classified within the ``Daily Charger'' profile, V2G participation had no statistically significant effect on net battery capacity ($\beta = 0.008 \times 10^{-4}, p = 0.966$; Appendix Figure \ref{fig:linear_regression_NMCGrb1}). Their vehicles tended to remain at high SOC for extended periods; as a result, although V2G participation increased capacity loss due to cycle aging ($\beta = -1.92 \times 10^{-4}, p = 1.24 \times 10^{-42}$), it substantially reduced calendar aging ($\beta = 1.93 \times 10^{-4}, p = 6.02 \times 10^{-9}$), effectively offsetting the added cycle degradation. For drivers classified within the ``BEV 2nd Vehicle'' profile, V2G participation led to slightly greater net capacity loss ($\beta = -0.87 \times 10^{-4}, p = 4.1 \times 10^{-5}$). Their vehicles tended to remain at mid-range SOC, so the reduction in calendar aging ($\beta = 0.97 \times 10^{-4}, p = 3.11 \times 10^{-16}$) was insufficient to offset the additional cycle aging incurred ($\beta = -1.84 \times 10^{-4}, p = 1.92 \times 10^{-15}$). However, the magnitude of capacity loss was less pronounced than that observed in the $\mathrm{LFP}\mid\mathrm{Gr}$ design. Finally, for drivers classified within the ``Threshold Charger'' profile, V2G participation led to a small but statistically significant improvement in capacity retention ($\beta = 2.40 \times 10^{-4}, p = 3.69 \times 10^{-12}$). In this case, the reduction in calendar aging ($\beta = 3.83 \times 10^{-4}, p = 2.24 \times 10^{-31}$) exceeded the added cycle aging ($\beta = -1.434 \times 10^{-4}, p = 4.3 \times 10^{-30}$). Figure \ref{fig:v2g_battery_degradation_vehicle_clusters} summarizes the capacity loss by profile, highlighting that, on average, both drivers classified within the ``Daily Charger'' and ``Threshold Charger'' profiles experienced capacity improvements from V2G participation.

Lastly, results from the $\mathrm{NMC}\mid\mathrm{Gr~B2}$ battery design are mixed. Overall, V2G participation generally leads to greater capacity loss for all vehicle usage profiles (linear regression comparing the influence of V2G kWh/year on total capacity change: all $\beta < -1.55 \times 10^{-4}$, all $p < 0.001$; Appendix Figure \ref{fig:linear_regression_NMCGrb2}). Similar to the $\mathrm{NMC}\mid\mathrm{Gr~B1}$ design, degradation in the $\mathrm{NMC}\mid\mathrm{Gr~B2}$ design is dominated by calendar aging, although to a lesser extent (Section \ref{subsec:no_V2G_simulations}). Consequently, although our V2G strategy reduces calendar aging, the benefit is less pronounced than with the $\mathrm{NMC}\mid\mathrm{Gr~B1}$ battery. For most drivers, this reduction is insufficient to offset the additional cycle aging from V2G. However, the magnitude of this increased loss is consistently smaller than that observed with the cycle-aging-dominant $\mathrm{LFP}\mid\mathrm{Gr}$ design. For instance, the added capacity loss for drivers classified within the ``Daily Charger'' profile with the $\mathrm{NMC}\mid\mathrm{Gr~B2}$ design was roughly half that of the same driver with the $\mathrm{LFP}\mid\mathrm{Gr}$ design ($\beta = -2.7 \times 10^{-4}$ vs. $\beta = -5.5 \times 10^{-4}$, respectively). Figure \ref{fig:v2g_battery_degradation_vehicle_clusters} summarizes the capacity loss by profile, highlighting that with $\mathrm{NMC}\mid\mathrm{Gr~B2}$, drivers classified within each vehicle usage profile would experience less capacity loss than with $\mathrm{LFP}\mid\mathrm{Gr}$.

\section{Methods}
\subsection{Vehicle dataset}
\label{subsec:vehicle_dataset}

This study utilized data from 315 Toyota bZ4X battery electric vehicles based in California from January 1, 2023, to April 30, 2023. Data was obtained from the vehicles’ controller area network (CAN) bus, from which we used the time-series measurements of the battery pack’s state of charge (SOC) and temperature for our simulations. We estimated time-series C-rate values by calculating the changes in successive SOC measurements over time and dividing those values by the nominal capacity of the vehicle’s battery pack (71.4 kWh) and used these estimations for our simulations. 

Data was actively recorded only when the vehicle was turned on, capturing driving activity but excluding periods of parking or charging. Charging events were identified at times when there was a measurable increase in SOC between the end of one trip (ignition off) and the start of the next (ignition on). For each event, we estimated the average power between trips by dividing the change in energy by the change in time between trips. This estimated power was used to infer the charging speed: if the power exceeded 19.2 kW (the nominal upper limit for AC Level-2 charging in the US~\cite{USDOT_2025_ChargerSpeeds}), the event was classified as DC Fast Charging (DCFC); otherwise, it was classified as AC Level-2 charging. For the DCFC events, we set the power output for our simulations to a constant 100 kW, the average of 50 kW (the U.S. lower bound for DCFC~\cite{USDOT_2025_ChargerSpeeds}) and 150 kW (Toyota bZ4X's maximum DC charging speed). For the AC Level-2 charging events, we set the power output to a constant 9.6 kW, a middle V2G charging/discharge speed scenario assumed by Southern California Edison (SCE; more on V2G scenario in section~\ref{subsec:V2G_strategy}). For all charging events, we assumed that charging began immediately upon vehicle plug-in and ceased once the vehicle reached its target SOC observed at the start of its next trip.

To model vehicle usage over a long ownership period, the empirical driving and charging profiles from the four-month period were cyclically repeated to span 15 years. This method relies on the simplifying assumption that driver behavior remains consistent over the vehicle's lifespan. We acknowledge that this approach does not account for long-term behavioral shifts or seasonal variations in driving patterns, charging behavior, and ambient temperature outside the observed January-to-April timeframe.

\subsection{Vehicle Usage Behavior Clustering}
\label{subsec:V2G_clustering}

We identified driver ``profiles'' from the real-world BEV data to assess the influence of different driving/charging behavior on V2G feasibility and battery degradation. To derive driver profiles we first extracted key behavioral features from time-series vehicle usage data that influence battery degradation and V2G feasibility. The behavioral features affecting battery degradation include the SOC at the start and end of charging, charging frequency, and charging duration. The behavior features affecting V2G feasibility include home charging availability and the times of day drivers choose to charge their vehicles. Appendix \ref{sec: list_behavioral_features} outlines the full list of behavioral features used for clustering. From the complete list of features, we performed feature selection via a Pearson correlation matrix, systematically removing highly correlated features to minimize bias in the k-means clustering process and enhance the interpretability of the resulting clusters. We then conducted k-means clustering~\cite{k_means_citation} iteratively with different numbers of clusters. We identified the optimal number of clusters using the elbow method, which identifies the point at which adding more clusters yields minimal reduction in the within-cluster sum of squares. This process identified four distinct profiles of vehicle usage behavior in our dataset, which are described in the results. 

\subsection{Battery degradation modeling}
\label{subsec:battery_model}
Our battery degradation simulations are based on adapted versions of semi-empirical models developed by Gasper et al. ~\cite{gasper2023degradation}. Their models predict capacity loss as the sum of calendar aging and cycle aging in large-format lithium-ion batteries. Calendar aging is dependent on time, SOC, and temperature, whereas cycle aging is dependent on the number of cycles, depth of discharge, C-rate, and temperature (Appendix \ref{sec: dynamic_battery model}, Equation \ref{first_equation}). Gasper et al. fitted the model parameters to accelerated aging test data from four commercial large-format lithium-ion batteries that varied in chemistry, design, and capacity. In this study, we adopt the published parameter sets for three of those batteries—$\mathrm{LFP}\mid\mathrm{Gr}$, $\mathrm{NMC}\mid\mathrm{Gr~B1}$, and $\mathrm{NMC}\mid\mathrm{Gr~B2}$ (the latter two differ in design and capacity). These batteries were not drawn directly from production battery-electric vehicles; however, their design and chemistries are representative of batteries that can be considered for vehicle use. 

The original semi-empirical models developed by Gasper et al. were calibrated with the static aging conditions of accelerated aging tests. To accommodate real-world dynamic vehicle usage, we reformulated these models into differential forms that could be solved iteratively, calculating incremental capacity loss at each time-step based on the vehicle’s instantaneous state of charge and C-rate (See Appendix \ref{sec: dynamic_battery model}). Given the mild climate variability in California, we assumed a constant operating temperature for each vehicle; specifically, the temperature value input into the model for each vehicle was the vehicle's average battery pack temperature recorded over the four-month data collection period. In addition, we assumed a constant graphite-to-reference potential ($U_a$). $U_a$ is a function of SOC, and our vehicle usage data shows that vehicles operated primarily in a mid-to-high SOC range, where there is a low variability in $U_a$. Appendix \ref{sec: dynamic_battery model}  Equations \ref{2_equation} - \ref{third_equation} outline the final dynamic forms of the models. Ultimately, given a vehicle's usage profile, the resulting models output the capacity loss attributed separately to calendar aging and cycle aging $\mathrm{LFP}\mid\mathrm{Gr}$, $\mathrm{NMC}\mid\mathrm{Gr~B1}$, and $\mathrm{NMC}\mid\mathrm{Gr~B2}$ over the vehicle’s usage period.

\subsection{Data Analysis}
\subsubsection{For Section \ref{subsec:no_V2G_simulations} Simulations of Battery Degradation under Californian Vehicle Usage}
From the simulated battery degradation outputs, we assessed whether a battery design's aging was primarily driven by calendar or cycle aging. We first calculated the ratio of calendar aging to total capacity loss for each individual vehicle simulation. We then analyzed the statistical properties of the resulting distribution of these ratios across the entire vehicle population. A Shapiro-Wilk test \cite{Shapiro_wilk} indicated that the ratio of calendar aging to total capacity loss was not normally distributed for each battery design ($\mathrm{LFP}\mid\mathrm{Gr}$: p = 0.001; $\mathrm{NMC}\mid\mathrm{Gr~B1}$: p = 0.01; $\mathrm{NMC}\mid\mathrm{Gr~B2}$: p = 0.003). As such, we applied Wilcoxon signed-rank tests~\cite{WILCOXON} to evaluate whether the median value of the calendar aging fraction significantly differed from 0.5 (0.5 represents a 50/50 contribution from both calendar and cycle aging) for each battery design. A statistically significant result ($p<0.05$) would confirm that the battery's degradation is not balanced between calendar aging and cycle aging; a median ratio significantly greater than 0.5 would indicate calendar aging dominance, whereas a median ratio significantly less than 0.5 would indicate cycle aging dominance.

We then evaluated whether a specific vehicle usage behavior led to statistically higher battery degradation for a given battery design. To do this, for each battery design, we first assessed the normality of the capacity loss distributions for each vehicle usage behavior cluster using the Shapiro-Wilk test. For all three battery designs, at least one cluster exhibited non-normality in its capacity loss values. Given this, we used the non-parametric Kruskal-Wallis test to evaluate whether the distributions of capacity loss differed significantly among the clusters. When this test indicated a significant difference existed, pairwise Mann-Whitney U tests were subsequently conducted to determine which specific clusters differed and to identify which cluster tended to exhibit higher capacity loss than the others.

\subsubsection{For Section \ref{subsec:V2G_simulations} Simulations of Incorporating V2G Strategy to California Vehicle Usage Profiles}

To determine the relationship between V2G participation and battery degradation across battery designs and driver behavior profiles, 
we performed separate linear regression analyses for each combination of battery design and vehicle usage behavior cluster. These analyses compared the impact of V2G participation (independent variable) on changes in battery capacity (dependent variable). V2G participation was quantified as the total energy dispatched to the grid per year (kWh/year). Change in capacity from V2G participation was quantified as the difference in battery capacity after 10 years of vehicle usage with V2G participation and the capacity after 10 years of vehicle usage without V2G participation. Statistical significance was evaluated using p-values: coefficients with p-values $<$ 0.05 are considered statistically significant, indicating that there is a relationship between V2G participation and degradation. Once statistical significance is established, we interpret the sign and magnitude of the coefficients to determine the direction and strength of the effect.

\section{Discussion}\label{sec12}

Our analysis resolves key uncertainties hindering the widespread adoption of V2G by leveraging real-world BEV usage data from California to design a user-centric V2G strategy and demonstrate its feasibility and effects on battery degradation. Our results show that the potential for drivers to participate in our V2G strategy is highly dependent on their BEV usage behavior. Furthermore, the resulting impact on battery degradation, a critical factor in the overall viability for the consumer, depends on the battery design’s sensitivity to calendar aging and the extent to which it is amplified by a driver’s vehicle usage behavior.

In quantifying the potential participation rates of BEV drivers under our user-centric V2G strategy, we find that drivers classified within the ``Daily Charger'' profile, representing 26\% of drivers in our California vehicle dataset, have the highest potential V2G participation, providing 1,259 kWh/yr on average. Conversely, drivers classified within the ``Public Charger'' profile, who represent 19\% of drivers, have the lowest potential, primarily because their reliance on public infrastructure prevents consistent overnight charging, which is necessary for our V2G strategy. We additionally found that drivers classified within the ``Threshold Charger'' and ``BEV 2nd Vehicle'' profiles, representing 38\% and 17\% of our dataset respectively, had lower potential V2G participation rates than the ``Daily Charger'' group because they did not plug in their vehicles overnight as frequently. However, these participation rates are derived from vehicle usage behaviors in 2023, when drivers were not exposed to specific incentives for overnight charging. It is possible that the deployment of our V2G strategy, which includes economic incentives for overnight charging, could drive a behavioral shift toward increasing the frequency of plugging in the vehicle overnight. Thus, our reported participation rates for these profiles should be interpreted as a conservative lower bound.



Our simulation analyses of the impact of V2G on battery longevity help explain some of the conflicting battery degradation results reported in existing literature. Our results suggest that for cycle-aging-dominant batteries such as the $\mathrm{LFP}\mid\mathrm{Gr}$ battery design, V2G participation is a straightforward cost; each additional cycle from V2G leads to further capacity loss across all vehicle usage behaviors (Figure~\ref{fig:v2g_battery_degradation_vehicle_clusters}). In contrast, for calendar-aging-dominant designs, such as the $\mathrm{NMC}\mid\mathrm{Gr~B1}$ battery, our simulations suggest that V2G participation can lead to negligible changes in capacity or even result in improved capacity retention. This counter-intuitive benefit arises because our overnight V2G strategy enables the battery to spend more time resting at a lower SOC, thereby reducing calendar aging to a degree that offsets or exceeds the additional cycle aging from V2G participation. This positive impact is most pronounced for ``Daily Charger,'' whose usage behavior, without V2G, inherently tends to keep their batteries at a high SOC. Ultimately, our study helps explain why there have been differences in results in existing literature and illustrates how these differences are attributable to the interplay of factors affecting battery degradation, including variations in battery designs and usage behaviors.

We note that in our simulations for the battery with the highest sensitivity to cycle aging ($\mathrm{LFP}\mid\mathrm{Gr}$), the total estimated additional degradation from V2G is relatively low, with an average 0.7\% capacity loss over 10 years for the vehicle cluster with the highest V2G usage (``Daily Charger''). Although our results may not generalize to all battery designs, they do suggest that under more real-world driving and charging constraints, the additional impact of V2G on battery degradation may not be as severe as consumers perceive and may be within ranges that they are willing to accept.

Our findings provide direct implications for BEV owners, OEMs and utilities. For BEV owners, our results resolve uncertainties related to battery degradation, vehicle usability, and V2G profits. By demonstrating that V2G does not accelerate degradation in all cases (and may, in fact, mitigate degradation for specific battery designs and usage behaviors), we challenge the perception that V2G is solely detrimental to battery health. Furthermore, our strategy ensures minimal disruption and sufficient driving range by restricting discharge to overnight idle periods and enforcing a 50\% SOC floor. Finally, by quantifying the financial potential for each driver profile, we provide clear expectations regarding V2G profitability across different usage behaviors. Collectively, these insights enable owners to evaluate the personal cost-benefit trade-off of V2G based on their vehicle and habits, potentially incentivizing future actions such as the installation of a home bidirectional charger to gain access to V2G.

For OEMs, our study demonstrates the importance of understanding both vehicle usage behavior and battery degradation when designing V2G schedules. OEMs are best suited to design and/or propose personalized V2G schedules that fit best with people's individual vehicle usage habits and could leverage this information to help consumers feel more confident in adopting V2G. Vehicle-based interfaces (e.g., in-cabin head units, vehicle apps) are drivers' main mode of controlling charging (e.g., charge schedulers) and monitor aspects of the vehicle (e.g., battery state of charge). Well designed interfaces could help mitigate consumer hesitancy around V2G, particularly if these provide information about potential compensation, battery degradation, and offer personalized control over charging~\cite{Monica_AutoUI, goncearuc2024barriers}. Consumers are interested in personalized schedules and suggestions that help them maximize monetary benefits while minimizing disruptions to their lives and the impact on their battery~\cite{Monica_AutoUI}. By leveraging individual driving patterns in ways similar to our V2G strategy, and with a clear understanding of the cycle and calendar aging profiles of the traction batteries in their vehicles, OEMs are well positioned to incorporate personalized V2G scheduling features in their existing charge management services and interfaces, which could help consumers feel confident in the benefits V2G could offer while also providing transparent control over their participation.


For utilities, our results provide critical metrics to assess V2G potential grid support and to structure fair and effective V2G compensation rates. With our study's estimated average annual energy contribution of each driver profile, ranging from 1259 kWh/yr for ``Daily Charger'' to just 111 kWh/yr for ``Public Charger'', and relative population proportion of each profile, utilities can forecast the potential aggregate load capacity available from our designed V2G strategy. Furthermore, with the estimated battery degradation under our V2G strategy, utilities can design V2G compensation rates that fairly account for the degradation cost to BEV owners.

The scope of our study is subject to three key limitations.

First, our findings are derived from Californian BEV data over a four-month period (January - April). While this dataset reflects real-world BEV usage within California, it may not be representative of BEV fleets in other regions, which may exhibit different climatic conditions and vehicle usage behaviors. Temperature is also known to affect calendar and cycle aging rates~\cite{Naumann2018, Smith2021, Dubarry2018}, and thus temperature variations could alter the degradation impact of any given V2G strategy~\cite{Hamid_V2G, Guo2019, Thingvad2021}. For instance, Movahedi et al. have shown that higher temperatures can increase a battery's sensitivity to calendar aging, which in turn can reduce the degradation impact from V2G participation~\cite{Hamid_V2G}. Therefore, the specific degradation impacts and V2G trade-offs we identified in our Californian dataset may not be directly applicable to regions with different climates. Furthermore, the four driver profiles identified and their relative proportions are specific to our dataset of BEVs in California and may not be representative of behaviors in other regions or of behaviors during different seasons. Since different vehicle usage behaviors would affect participation rates and the corresponding degradation, the potential and viability of our V2G strategy can change with different driving cultures or with different seasons.

Second, our results are contingent upon the specific parameters of our user-centric V2G strategy defined in Section \ref{subsec:V2G_strategy}. Alternative V2G strategies could yield different V2G compensations, participation rates, and degradation outcomes. For example, a strategy that discharges BEVs at a different time of day (e.g. 9 pm - 12 am instead of 6 pm - 9 pm) would alter compensation, as V2G compensation rates, such as those proposed by Southern California Edison, change with the time of day~\cite{SCE_V2G_pilot_program}. Similarly, a strategy allowing more V2G discharging events could raise participation rates but simultaneously accelerate degradation~\cite{Dubarry2017, Hamid_V2G}. Furthermore, a strategy that discharges to a lower SOC level (i.e., to $40\%$ SOC instead of $50\%$) might further improve the degradation for calendar-aging dominant batteries~\cite{Wei2022}, but could also deter drivers from V2G participation. Ultimately, different V2G strategies alter the cost-benefit trade-off of V2G for the BEV owner.

Third, the fidelity of our results relies on the semi-empirical degradation models detailed in Section \ref{subsec:battery_model}. These models are adapted from the semi-empirical framework developed by Gasper et al. \cite{gasper2023degradation}, which is calibrated on static aging conditions from accelerated aging tests. Our adaptation converts these models into a state-dependent form, enabling the use of dynamic real-world usage profiles as input. However, this adaptation assumes that degradation under dynamic loads is the cumulative sum of static aging steps, a simplification that ignores potential path-dependent degradation mechanisms. Consequently, this assumption can affect the accuracy of the estimated capacity loss values. Our simulations are most valuable for illustrating the relative trade-offs and trends of V2G participation across different battery designs and driver profiles, rather than establishing exact transferable predictions.


Widespread adoption of V2G is hindered by stakeholder uncertainties regarding its potential effects on battery degradation and vehicle usability. This study resolves these uncertainties by leveraging real-world usage data from EVs in California to first design a user-centric V2G strategy and then simulate its effects across different vehicle usage behaviors and battery designs.

From the real-world usage dataset of EVs in California, we distill BEV usage profiles into four representative driving profiles critical for V2G analysis. Our analysis suggests that participation potential in our user-centric overnight V2G strategy is highly dependent on vehicle usage behavior. ``Daily Chargers'' (26\% of the population) have the highest V2G participation potential, contributing an average of 1,259 kWh/year to the grid, while ``Public Chargers'' (19\% of the population) have the lowest as they primarily fast charge from public chargers. Furthermore, our results indicate the effect of V2G participation on battery degradation depends on the battery design’s sensitivity to calendar aging and the extent to which a vehicle usage behavior amplifies calendar aging. For battery designs with low sensitivity to calendar aging, such as the cycle-aging-dominant $\mathrm{LFP}\mid\mathrm{Gr}$ battery, V2G participation invariably increases capacity loss for all vehicle usage behaviors due to additional cycle aging. However, for battery designs where calendar aging is the dominant degradation mode, such as the $\mathrm{NMC}\mid\mathrm{Gr~B1}$ battery, participating in our user-centric V2G strategy can lead to negligible changes in capacity loss or even improved capacity retention, particularly for drivers like those in the ``Daily Charger'' profile who typically keep their batteries at high SOC. Our V2G strategy reduces the time the battery spends at a high state of charge, which for calendar-aging-dominant batteries, reduces calendar aging enough to offset or even exceed the additional cycle aging.

Our findings directly address the stakeholder uncertainties hindering V2G adoption. These results clearly inform BEV owners about the costs and benefits of V2G participation specific to their personal vehicle usage behaviors and vehicle battery type. Furthermore, our analysis quantifies for utilities the potential grid support from drivers participating in our designed V2G strategy and enables utilities to design V2G compensation rates that fairly account for the degradation cost. Lastly, our results demonstrate to OEMs that V2G can be an appealing feature that can potentially prolong the lifetime of BEV batteries as well as generate revenue for BEV owners rather than solely be a liability.

Taken together, our study provides the scientific basis necessary to clearly inform stakeholders about the financial benefits and trade-offs of V2G. This work enables future field studies, in which we will present personalized financial and battery health impacts directly to BEV owners, empowering them with clear insight into the viability of V2G participation~\cite{clement_autoui}.

\backmatter





\section*{Declarations}


\begin{itemize}
    \item \textbf{Funding} \\
    This work was supported by the Toyota Research Institute.
    \item \textbf{Conflict of interest/Competing interests} \\
    The authors declare no competing interests.
    \item \textbf{Data availability}\\
    Data generated by simulations and used in our analyses will be made publicly available upon final publication.
    \item \textbf{Code availability} \\
    Code used to generate simulations, figures, and reported analyses will be made publicly available upon final publication.
    \item \textbf{Author contribution} \\
    \textbf{Clement Wong}: Conceptualization, Methodology, Software, Validation, Formal analysis, Investigation, Writing – original draft, Writing – review \& editing, Visualization. 
    \textbf{Amalie E. Trewartha}: Conceptualization, Methodology, Writing – Review \& Editing, Supervision.
    \textbf{Steven B. Torrisi}: Conceptualization, Methodology, Writing – Review \& Editing, Supervision.
    \textbf{Alexandre L.S. Filipowicz}: Conceptualization, Methodology, Formal analysis, Investigation, Resources, Data Curation, Writing – review \& editing, Supervision.
\end{itemize}







\clearpage

\begin{appendices}
\section{Transforming battery model to be dynamic}
\label{sec: dynamic_battery model}

Gasper et al. proposed a semi-empirical model, outlined in Equation \eqref{first_equation}, to predict the capacity for four commercial large-format lithium-ion batteries ($\text{LFP}|\text{Gr}$, $\text{NMC}|\text{Gr~A}$, $\text{NMC}|\text{Gr~B1}$, and $\text{NMC}|\text{Gr~B2}$). The four batteries varied in chemistry, design, and capacity (see \cite{gasper2023degradation} for specifications). Their model separates the total capacity loss into two components: calendar aging and cycle aging.

\begin{equation}
\label{first_equation}
\begin{aligned}
q &= q_0 - q_{\text{Loss}}, \qquad q_{\text{Loss}} = q_{\text{Loss,Cal}} + q_{\text{Loss,Cyc}} \\
q_{\text{Loss,Cal}} &= p_1 \cdot \exp\left(\frac{p_2}{T}\right) \cdot \exp\left(\frac{p_3 U_a (\text{SOC})}{T}\right) \cdot t^{p_4} \\
U_a(\text{SOC}) &= 0.6379 + 0.5416 \exp\left(-305.53 \cdot (0.0085 + 0.7715 \cdot \text{SOC})\right) \\
&\quad + 0.044 \tanh\left(\frac{-(0.0085 + 0.7715 \cdot \text{SOC}) - 0.1958}{0.1088} \right) \\
&\quad - 0.1978 \tanh\left(\frac{(0.0085 + 0.7715 \cdot \text{SOC}) - 1.0571}{0.0854} \right) \\
&\quad - 0.6875 \tanh\left(\frac{(0.0085 + 0.7715 \cdot \text{SOC}) + 0.0117}{0.0529} \right) \\
&\quad - 0.0175 \tanh\left(\frac{(0.0085 + 0.7715 \cdot \text{SOC}) - 0.5692}{0.0875} \right)\\
q_{\text{Loss,Cyc}} &= \left(p_5 + p_6 \cdot \text{DOD} + p_7 \cdot C_{\text{rate}}\right) \cdot \left(\exp\left(\frac{p_8}{T}\right) + \exp\left(\frac{-p_9}{T}\right)\right) \cdot \text{EFC}^{p_{10}}
\end{aligned}
\end{equation}

Here, $q$ is the remaining capacity, $q_0$ is the initial capacity, and $q_{\text{Loss}}$ is the total capacity loss, composed of calendar aging ($q_{\text{Loss,Cal}}$) and cycle aging ($q_{\text{Loss,Cyc}}$). Calendar aging is a function of time ($t$), temperature ($T$), and graphite-to-reference potential ($U_a$), which depends on the state of charge (SOC). Cycle aging is a function of depth of discharge (DOD), C-rate, temperature ($T$), and the total number of equivalent full cycles (EFC). Parameters $p_1, \dots, p_{10}$ are fitted to accelerated calendar- and cycle-aging test data for the four batteries. 

Gasper's model (Equation \ref{first_equation}) can only be implemented over standard testing cycles (i.e., fully charging/discharging profiles with fixed current). To make the model applicable to real-world electric vehicle usage, we reformulated it to accept time-varying C-rate and SOC inputs as shown in Equation \ref{2_equation}. The expressions for DOD and EFC were reformulated as functions of SOC and C-rate ($\text{EFC}= \int_0^t C_\text{rate} (\tau)d\tau$, $\text{DOD}= 1-\text{SOC}$).

\begin{equation}
\label{2_equation}
\begin{aligned}
q(t) &= q_0 - \left( q_{\text{Loss,Cal}}(t) + q_{\text{Loss,Cyc}}(t) \right) \\
q_{\text{Loss,Cal}}(t) &= p_1 \cdot \exp\left(\frac{p_2}{T(t)}\right) \cdot \exp\left(\frac{p_3 U_a(\text{SOC}(t))}{T(t)}\right) \cdot t^{p_4}\\
q_{\text{Loss,Cyc}}(t) &= \left(p_5 + p_6 \cdot \left(1-\text{SOC}(t)\right) + p_7 \cdot C_{\text{rate}}(t)\right) \\
&\quad \cdot \left(\exp\left(\frac{p_8}{T(t)}\right) + \exp\left(\frac{-p_9}{T(t)}\right)\right) \cdot \left(\int_{0}^{t} C_{\text{rate}}(\tau) \, d\tau\right)^{p_{10}}
\end{aligned}
\end{equation}

For each vehicle, we assumed temperature to be constant over time, given the relatively mild climate variability in California. Accordingly, temperature was approximated as the time-averaged battery pack temperature over the four-month period:
\begin{equation}
\label{eq:time_averaged_temperature}
T(t) \approx T_{\mathrm{avg}} = \frac{1}{t_N - t_0} \sum_{i=0}^{N-1} \frac{ T(t_i) + T(t_{i+1}) }{2} \left( t_{i+1} - t_i \right),
\end{equation}
where $t_0$ and $t_N$ denote the initial and final instances of times, respectively.

In addition, for each vehicle, we assumed the graphite-to-reference potential to be constant over time, as our vehicle-usage data indicate that the majority of vehicles operated within a SOC range over which $U_a$ exhibits low variability. Accordingly, the graphite-to-reference potential was approximated as the value corresponding to the time-averaged SOC over the four-month observation period:
\begin{equation}
\label{eq:time_averaged_graphite_potential}
\begin{aligned}
U_a(\text{SOC}(t)) &\approx U_a(\text{SOC}_{\mathrm{avg}}) \\
\text{where } \quad \text{SOC}_{\mathrm{avg}} &= \frac{1}{t_N - t_0} \sum_{i=0}^{N-1} \frac{ \text{SOC}(t_i) + \text{SOC}(t_{i+1}) }{2} \left( t_{i+1} - t_i \right).
\end{aligned}
\end{equation}

By substituting Equations \ref{eq:time_averaged_temperature} and \ref{eq:time_averaged_graphite_potential} into Equation \ref{2_equation} and differentiating with respect to time, a dynamic version of the semi-empirical model can be expressed as:

\begin{equation}
\label{third_equation}
\begin{aligned}
q(t+\Delta t) &= q_0 -  (q_{\text{Loss,Cal}}(t+\Delta t) + q_{\text{Loss,Cyc}}(t+\Delta t)) \\
q_{\text{Loss,Cal}}(t+\Delta t) &= p_1 \cdot \exp\left(\frac{p_2}{T_\mathrm{avg}}\right) \cdot \exp\left(\frac{p_3 U_a(SOC_\mathrm{avg})}{T_\mathrm{avg}}\right) \cdot (t+\Delta t)^{p_4 }\\
q_{\text{Loss,Cyc}}(t+\Delta t) &= q_\text{Loss,Cyc}(t) + \Delta t \cdot \frac{dq_{\text{Loss,Cyc}}}{dt}(t)
\\
\frac{dq_{\text{Loss,Cyc}}}{dt} (t)&= \left((p_5 + p_6 \cdot (1 - \text{SOC}(t))) + p_7 \cdot C_{\text{rate}}(t)\right) \\&\cdot \left(\exp\left(\frac{p_8}{T_\mathrm{avg}}\right) + \exp\left(-\frac{p_9}{T_\mathrm{avg}}\right)\right) \cdot p_{10} \\&\cdot \left(\int_0^t C_{\text{rate}}(\tau) \, d\tau\right)^{p_{10}-1} \cdot C_{\text{rate}}(t)
\end{aligned}
\end{equation}
where $t$ is the instantaneous value of time, and $\Delta t$ is the sampling period. Using Equation \ref{third_equation}, we can simulate the capacity of $\text{LFP}|\text{Gr}$, $\text{NMC}|\text{Gr~B1}$, and $\text{NMC}|\text{Gr~B2}$ given a vehicle's usage in time-varying C-rate and SOC.

\section{Full list of Behavioral Features for Vehicle Usage Behavior Clustering }
\label{sec: list_behavioral_features}
\begin{multicols}{2}
\begin{itemize}
    \item Number of days the vehicle was driven
    \item Total trips logged
    \item Cumulative hours in active use
    \item Cumulative hours parked
    \item First date in the dataset
    \item Last date in the dataset
    \item Total charging sessions
    \item Number of slow charging sessions
    \item Number of DC-fast charging sessions
    \item Average charging power (kW)
    \item Charging sessions at the home location
    \item Average SOC before charging (\%)
    \item Average SOC after charging (\%)
    \item Time-weighted SOC consumed while driving (\%-h)
    \item Total hours in active use
    \item Time-weighted SOC while parked (\%-h)
    \item Total hours parked
    \item Average SOC during driving episodes (\%)
    \item Total SOC percentage points consumed by driving
    \item Total energy charged (kWh)
    \item Total calendar days observed
    \item Hours in use across all days
    \item Hours parked across all days
    \item Average SOC consumed per driving day (\%)
    \item Average daily SOC consumption across all days (\%)
    \item Average kWh charged per driving day
    \item Average daily charged energy across all days (kWh)
    \item Time-weighted mean SOC during use (\%)
    \item Time-weighted mean SOC while parked (\%)
    \item Average trips per driving day
    \item Average charging sessions per driving day
    \item Average AC-slow charges per driving day
    \item Average DC-fast charges per driving day
    \item Total trips recorded
    \item Behavioral cluster label
    \item Overall average state-of-charge (\%)
    \item Mean C-rate during charging
    \item Mean C-rate during discharge
    \item Equivalent full cycles accumulated from charging
    \item Equivalent full cycles accumulated from driving
\end{itemize}
\end{multicols}

\section{V2G preferences survey}
\label{sec: survey}

The minimum state-of-charge used in the V2G simulation was estimated based on a survey conducted by the Toyota Research Institute on people's attitudes and preferences towards V2G. The survey was conducted in June 2024 and administered to a representative sample of US-based new vehicle intenders, who were identified and recruited by the marketing firm Escalent.

The survey was administered to a representative sample of 1,753 participants from Escalent's annual EV Forward cohort who either owned, were actively looking to purchase, or would be open to considering a BEV (68\% of Escalent's EV Forward cohort). At the start of the survey, participants were first asked to rate their familiarity with V2G technology. Next, after a brief explanation of V2G, participants answered a series of questions related to their attitudes, motivators, and barriers towards V2G adoption. The survey took approximately 20 minutes to complete, and participants were paid \$14 for their time.

As a part of this survey, participants were asked the minimum number of miles of range (indicated as a number between 0 and 300 miles) they would need in their vehicle to participate in a V2G program (i.e., ``If you were to participate in a V2G program, what is the minimum miles of range you would always want available in your vehicle?''). Overall, the mean minimum range indicated was 187 miles (95\% CI: 181--192 miles), or 62\% of the 300 mile maximum range in the question. However, participants who owned a BEV (6\% of participants) indicated a lower mean minimum range of 157 miles (95\% CI: 141--174 miles), or 52\% of the 300-mile maximum range in the question. This difference aligns well with research finding that experience driving a BEV lowers range anxiety and perceived minimum mileage needs~\cite{rauh2015understanding, ma2025range}. As such, we chose a minimum SOC of 50\% for our V2G simulations, aligning more closely with the stated needs expressed by people with experience owning a BEV.

\section{Supporting Material for Simulations of Battery Degradation under Californian
Vehicle Usage}

\begin{table}[h!]
\centering
\caption{To assess if battery design is dominated by calendar aging or cycle aging. Wilcoxon signed-rank test for each battery design, where $H_0$: median (calendar aging)/(total capacity loss) = 0.5, and $H_1$: median (calendar aging)/(total capacity loss) $\ne$ 0.5}
\renewcommand{\arraystretch}{1.3}
\begin{tabular}{lcccc}
\toprule
 & \makecell{\textbf{Median} \\ \textbf{(calendar aging)} \\ $\boldsymbol{/(}$\textbf{total capacity loss}$\boldsymbol{)}$}
 & \makecell{\textbf{95\% CI} \\ \textbf{lower limit}} 
 & \makecell{\textbf{95\% CI} \\ \textbf{upper limit}} 
 & \textbf{p value} \\
\midrule
LFP$\mid$Gr & 0.395 & 0.380 & 0.410 & 2.89E-28 \\
NMC$\mid$Gr B1 & 0.794 & 0.785 & 0.803 & 2.14E-53 \\
NMC$\mid$Gr B2 & 0.591 & 0.573 & 0.609 & 1.82E-17 \\
\bottomrule
\end{tabular}
\label{tab:calendar_vs_cycle}
\end{table}

\begin{table}[h!]
\centering
\caption{To support for LFP$\mid$Gr: \textit{``Public Charger'' have statistically significantly higher capacity loss than other vehicle usage behaviors} \\
Mann–Whitney U test \\
Let $\mu_P$ = mean capacity loss of ``Public Charger'', $\mu_O$ = mean capacity loss of other behavior. \\
$H_0$: $\mu_P \leq \mu_O$, \quad $H_1$: $\mu_P > \mu_O$}
\renewcommand{\arraystretch}{1.3}
\begin{tabular}{lcccccc}
\toprule
\textbf{$H_1$} & $n_P$ & $n_O$ & $\mu_P (\%)$ & $\mu_O (\%)$ & $U_{\text{stat}}$ & $p_{value}$\\
\midrule
$\mu_P > \mu_O$ (vs. ``Daily Charger'') & 55 & 83 & 8.46 & 7.89&	2566	&1.09E-01 \\
$\mu_P > \mu_O$ (vs. ``BEV 2nd Vehicle'') & 55 & 56 & 8.46 & 5.36&	2669&	1.41E-11 \\
$\mu_P > \mu_O$ (vs. ``Threshold Charger'') & 55 & 121 & 8.46 & 6.61	&4798	&1.35E-06 \\
\bottomrule
\end{tabular}
\label{tab:LFP|Gr_public_charger}
\end{table}

\begin{table}[h!]
\centering
\caption{To support for NMC$\mid$Gr B1: \textit{``Daily Charger'' have statistically significantly higher capacity loss than other vehicle usage behaviors} \\
Mann–Whitney U test \\
Let $\mu_D$ = mean capacity loss of ``Daily Charger'', $\mu_O$ = mean capacity loss of other behavior. \\
$H_0$: $\mu_D \leq \mu_O$, \quad $H_1$: $\mu_D > \mu_O$}
\renewcommand{\arraystretch}{1.3}
\begin{tabular}{lcccccc}
\toprule
\textbf{$H_1$} & $n_D$ & $n_O$ & $\mu_D (\%)$ & $\mu_O (\%)$ & $U_{\text{stat}}$ &$p_{value}$ \\
\midrule
$\mu_D > \mu_O$ (vs. ``Public Charger'') & 83 & 55 & 13.89 & 12.34 & 3654	& 1.24E-09 \\
$\mu_D > \mu_O$ (vs. ``BEV 2nd Vehicle'') & 83 & 56 & 13.89 & 12.77 & 3277 & 2.15E-05 \\
$\mu_D > \mu_O$ (vs. ``Threshold Charger'') & 83&	121	&13.89	&12.99	&6976	&1.19E-06 \\
\bottomrule
\end{tabular}
\label{tab:NMC|Grb1_daily_charger}
\end{table}

\begin{table}[h!]
\centering
\caption{To support for NMC$\mid$Gr B2: \textit{``No single vehicle usage behavior group shows statistically significantly higher capacity loss''} \\
Mann–Whitney U test \\
Let $\mu_P$ = mean capacity loss of ``Public Charger'', $\mu_O$ = mean capacity loss of other behavior. \\
$H_0$: $\mu_P \leq \mu_O$, \quad $H_1$: $\mu_P > \mu_O$}
\renewcommand{\arraystretch}{1.3}
\begin{tabular}{lcccccc}
\toprule
\textbf{$H_1$} & $n_P$ & $n_O$ & $\mu_P (\%)$ & $\mu_O (\%)$ & $U_{\text{stat}}$ & $p_{value}$ \\
\midrule
$\mu_P > \mu_O$ (vs. ``Daily Charger'') & 55 & 83 & 6.32 & 6.36 & 2197 & 0.356 \\
$\mu_P > \mu_O$ (vs. ``BEV 2nd Vehicle'') & 55 & 56 & 6.32 & 4.79 & 2356 & 7.55E-07 \\
$\mu_P > \mu_O$ (vs. ``Threshold Charger'') & 55 & 121 & 6.32 & 5.70 & 4029 & 0.0126 \\
\bottomrule
\end{tabular}
\label{tab:NMC|Grb2_public_charger}
\end{table}

\begin{figure} [h!]
    \centering
    \includegraphics[width=\linewidth]{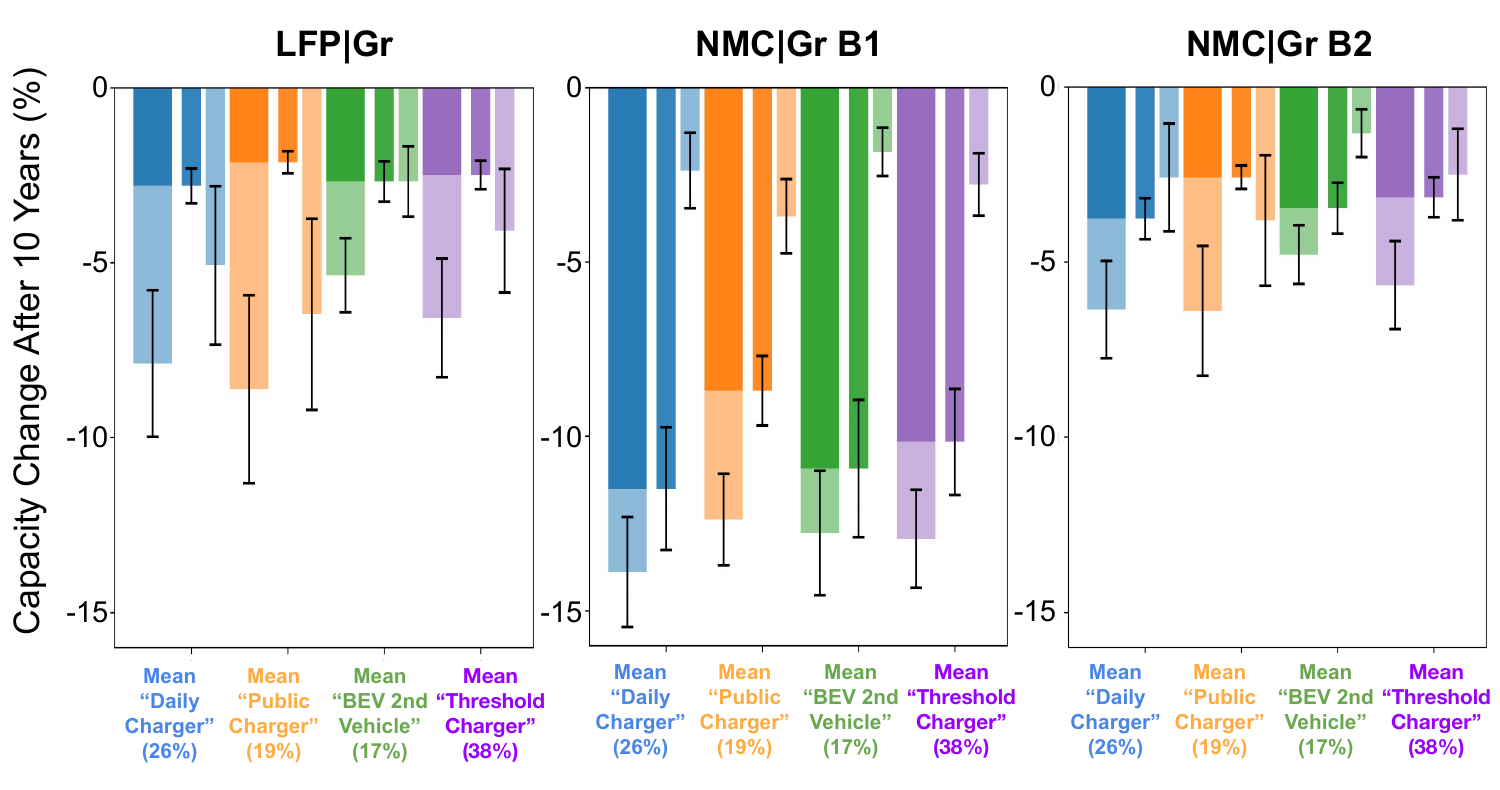}
    \caption{Simulations of the degradation of various battery designs under the usage of each California vehicle usage behavior cluster. Darker colors correspond to calendar aging while lighter colors correspond to cycle aging. Error bars indicate the 95\% CI population values.}
    \label{fig:battery_degradation_vehicle_clusters_with_error}
\end{figure}

\clearpage

\section{Linear Regression Analysis Supporting Results in Simulations of Incorporating V2G Strategy to California Vehicle Usage}

\begin{figure}[h!]
    \centering
    \includegraphics[width=1\linewidth]{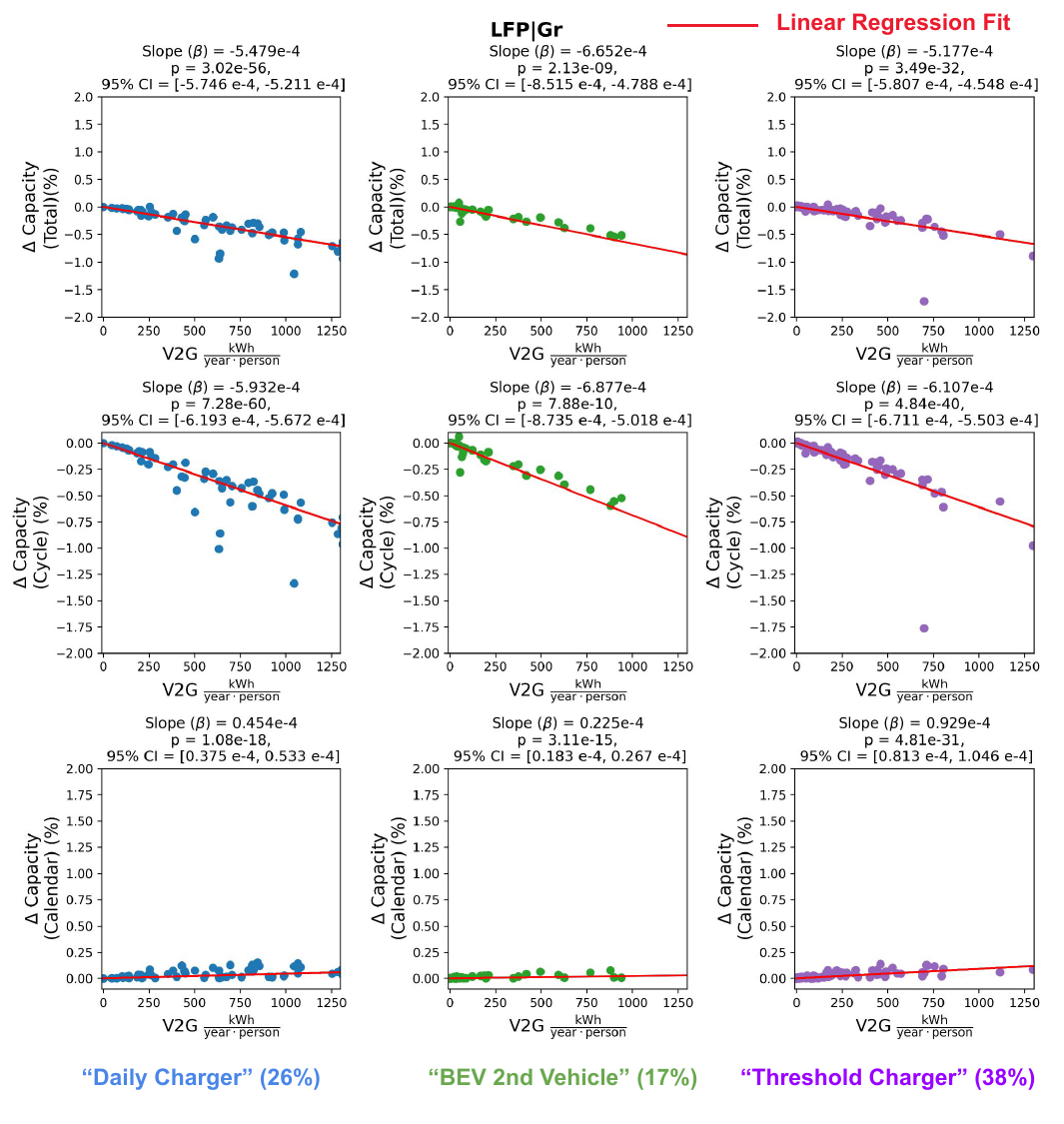}
    \caption{LFP$\mid$Gr linear regression comparing the influence of V2G kWh/year on change in total capacity, change in capacity attributed to cycle aging, change in capacity attributed to calendar aging}
    \label{fig:linear_regression_LFP}
\end{figure}

\begin{figure}[h!]
    \centering
    \includegraphics[width=1\linewidth]{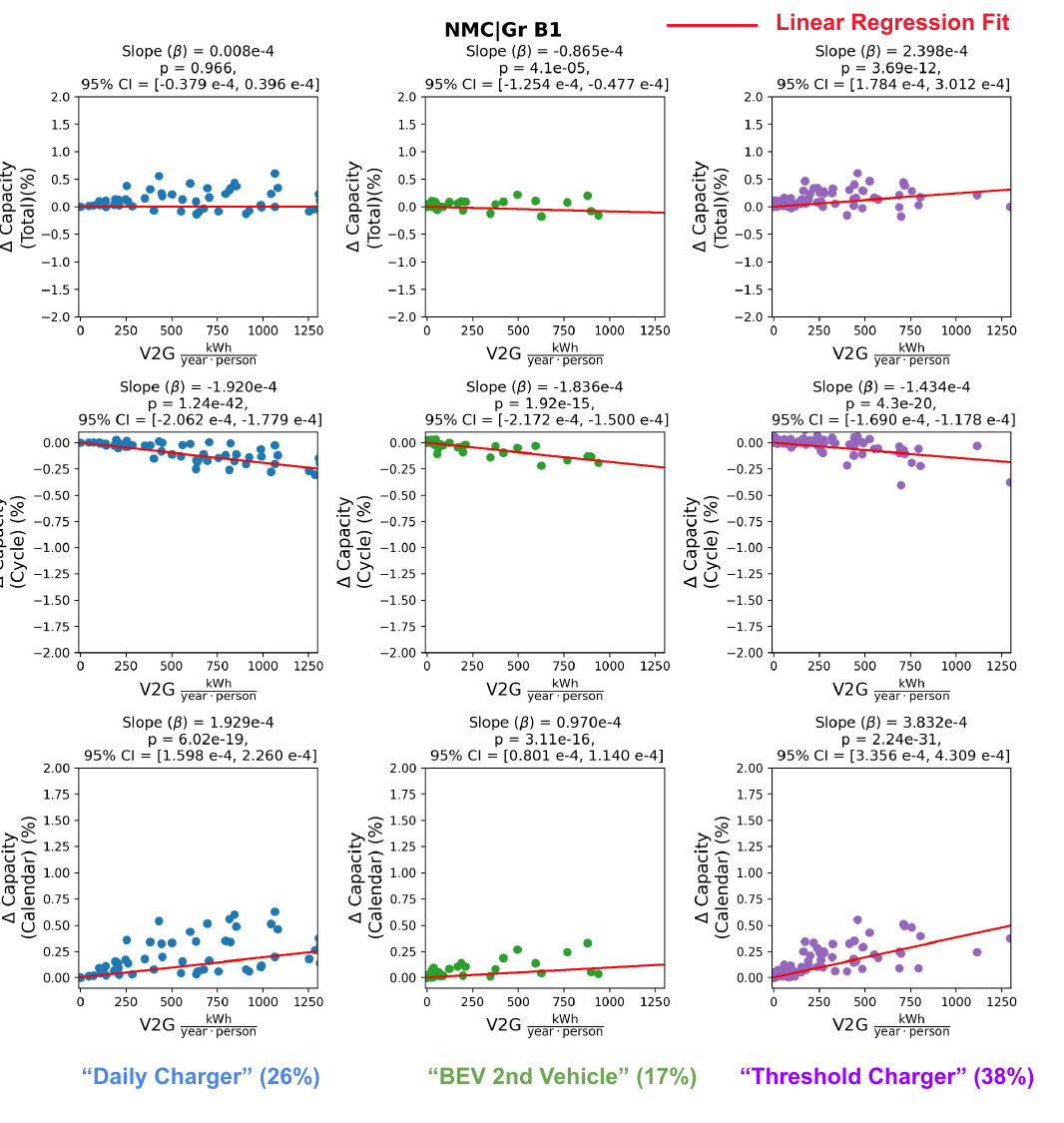}
    \caption{NMC$\mid$Gr B1 linear regression comparing the influence of V2G kWh/year on change in total capacity, change in capacity attributed to cycle aging, change in capacity attributed to calendar aging}
    \label{fig:linear_regression_NMCGrb1}
\end{figure}

\begin{figure}[h!]
    \centering
    \includegraphics[width=1\linewidth]{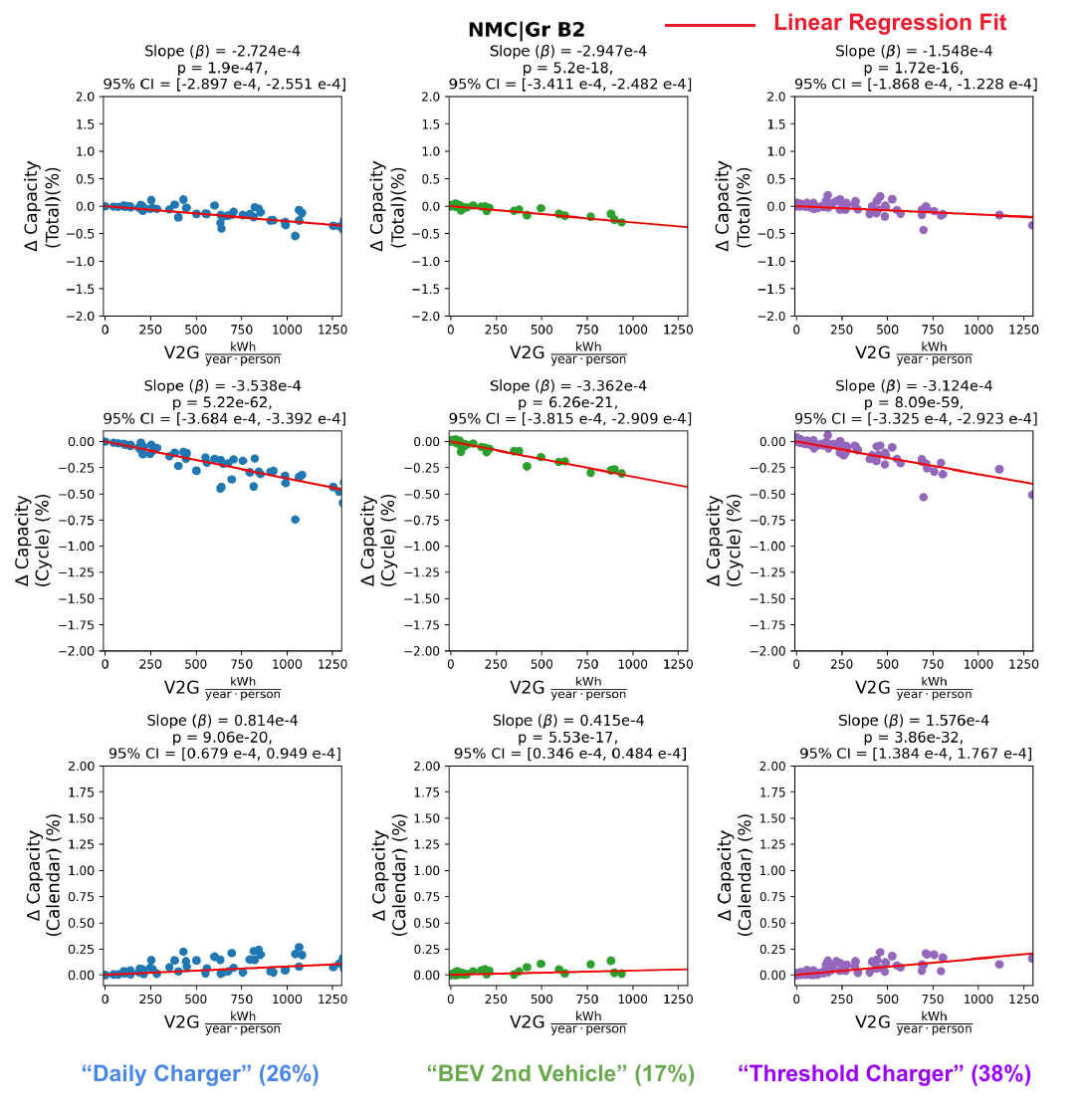}
    \caption{NMC$\mid$Gr B2 linear regression comparing the influence of V2G kWh/year on change in total capacity, change in capacity attributed to cycle aging, change in capacity attributed to calendar aging}
    \label{fig:linear_regression_NMCGrb2}
\end{figure}

\end{appendices}
\clearpage


\bibliography{bibliography_2-23}

\end{document}